\documentclass[aps,prl,groupedaddress,preprint,showpacs]{revtex4}

\usepackage[dvips]{graphicx}

\begin{document}

\title{Spectroscopy by frequency entangled photon pairs}

\author{Atsushi Yabushita}
 \author{Takayoshi Kobayashi}
 \affiliation{Core Research for Evolutional Science and
 Technology(CREST)}
\affiliation{Japan Science and Technology Corporation (JST)}
\affiliation{Department of Physics, Graduate School of Science, University of Tokyo, 7-3-1 Hongo, Bunkyo, Tokyo, 113-0033, Japan}

\date{\today}

\begin{abstract}
Quantum spectroscopy was performed using the frequency-entangled
 broadband photon pairs generated by spontaneous 
 parametric down-conversion. An absorptive
 sample was placed in front of the idler photon detector, and the
 frequency of signal photons was resolved by a diffraction
 grating. The absorption spectrum of the sample was
 measured by counting the coincidences, and the result is
 in agreement with the one measured by a conventional spectrophotometer
 with a classical light source.
\end{abstract}


\maketitle

\section{Introduction}

The polarization entanglement of spatially separated photon pairs,
generated by spontaneous parametric down-conversion (SPDC), has been
used in a variety of quantum experiments to demonstrate quantum
teleportation, entanglement-based
quantum cryptography, Bell-inequality
violations, and others~\cite{quantum-review}.

These SPDC photon pairs are also entangled in their wave vectors. The
entanglement in wave-vector space is used in various experiments, such
as quantum
imaging~\cite{quantum-imaging-lithography,quantum-imaging}, photonic de
Broglie wavelength
measurement~\cite{deBroglie,deBroglie2,deBroglie3}, quantum
interference~\cite{quantum-interference2,quantum-interference3},
 and quantum
lithography~\cite{quantum-lithography,quantum-lithography2,quantum-lithography3}.
It was claimed and experimentally verified that they have higher
resolution than the classical limit, which will be used to various
applications.

On the other hand, in the case of quantum imaging, the information
about the shape of a spatial filter is transferred by entangled photon pairs,
therefore it is useful for secure two-dimensional information
transfer. Compared with the case using the polarization entanglement, it
can send much more information by the entanglement of the wave-vector
space taking an advantage of the continuity of the entangled
parameter. A protocol for quantum key distribution was proposed based on
a system whose dimension is higher than 2 in Ref.\cite{QKD-highdimension}.

In this paper, a frequency entanglement was used to measure the
spectroscopic property of a sample, which can also be used for nonlocal pulse
shaping~\cite{nonlocal-pulseshaping}. The state was maximally entangled
by using a continuous-wave (cw)
pump~\cite{coincident-frequencies-theory,coincident-frequencies-theory2,coincident-frequencies-theory3}.
Focusing the pump on a SPDC crystal by an objective lens, the bandwidth
of SPDC was extended. It enabled to measure the absorption spectrum of a
sample in broadband. Various types of spectroscopic measurements were
performed utilizing classical light source including sophisticatedly
constructed extremely short pulses~\cite{classical-spectroscopy}. However,
in the following situations, spectroscopy utilizing the frequency
entanglement can be a powerful way to measure the spectroscopic
properties of the sample.

One of the situations is the case when it is difficult to analyze
frequency of photons transmitted through an absorptive sample. For
example, to measure the spectroscopic property of a sample in vacuum
ultraviolet (VUV) range, a spectrometer must be settled in a vacuum
chamber in case of the conventional VUV spectroscopy, and the
spectrometer must be aligned and controlled under the vacuum
condition. However, in case of spectroscopy using the
frequency-entangled photon pair consists of a VUV photon and a longer
wavelength
photon (like a visible photon), the VUV photon transmitted thorough the
sample is to be detected by the photodetector without capability of
energy resolution. Instead the visible photon is to be resolved by a
spectrometer in the atmospheric pressure which is easily handled. This
is one of the useful features of this method. The method is also useful
when the sample is in space and the photons transmitted through the
sample are not easy to be resolved by their energy.

Another situation is the case when the spectroscopic property of a
fragile sample is to be measured in infrared range. The power of the
light source must be very low not to damage the sample, and yet an
infrared photodetector is usually so noisy that a signal-to-noise ratio
is very low in the measurement using the classical light
source. However, when the frequency-entangled photon pairs are used, the
coincidence counts are measured, then one of the photons of each photon
pair works as a timing gate for the measurement of another photon of the
pair which is to be resolved by its energy, and the signal-to-noise
ratio is expected to be enhanced substantially. This is another
advantage of this method.

The SPDC photon pairs are emitted conically from the point where
the pump light is focused on the beta-barium borate (BBO) crystal. Using a parabolic
mirror, all SPDC photon pairs in the light cones are collimated
without achromatic aberration, and travel to a beam
splitter without expanding their beam diameters keeping their
polarization entanglement.
It is not aimed to use the polarization in this experiment, so signal
and idler photons are separated from each other by a polarizing beam
splitter which destroys the polarization entanglement.
When a nonpolarizing beam splitter is used in place of the
polarizing beam splitter, the photon pairs detected at the photon
counters have polarization entanglement, which can be used for quantum
cryptography~\cite{quantum-cryptography,B92}. Therefore using the
polarization entanglement and the frequency entanglement simultaneously,
it will become possible to implement wavelength-division multiplexing
(WDM), quantum cryptography, or WDM quantum key distribution, which can
send much more information compared with the case of polarization-only
entanglement.

\section{Experiment}

   \begin{figure}[htbp]
    \begin{center}
     \includegraphics[scale=0.4]{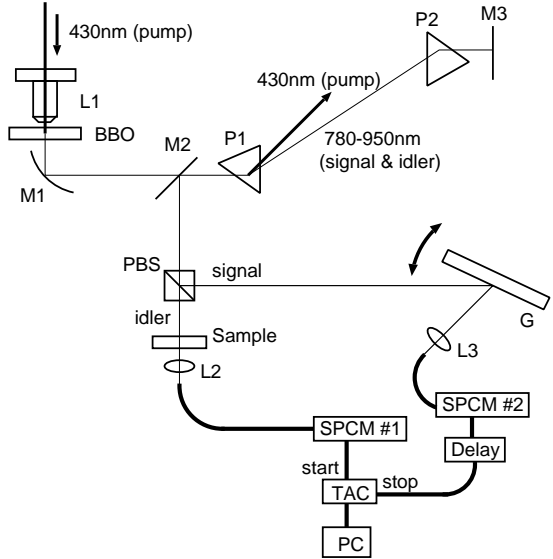}
    \end{center}
    \caption{Experimental setup. Details are given in the
    text.\label{figures-setup}}
   \end{figure}
The schematic drawing of our experimental setup is shown in
Fig.~\ref{figures-setup}.
Frequency-nondegenerate photon pairs are generated by
SPDC in a 1-mm-thick type-II BBO ($\beta$-BaB$_2$O$_4$) crystal pumped by the
second harmonic light (1.5~mW) of a cw Ti:sapphire laser operated at
859.4~nm. To generate photon pairs by bandwidth extended
SPDC, the pump light is focused on the BBO crystal by a microscope
objective lens
of 8-mm focal length (L1). Generated SPDC photons diversing
from the focal waist are
collimated by an off-axis parabolic mirror (M) of 25.4-mm focal
length. A prism pair (P1,P2) is used to eliminate the remainder of the
pump light, which can be a noise source in the experiments. When the
light beam passes through the prism pairs, the beam height was lowered by a
mirror (M3), and only the SPDC pairs are picked out by another mirror
(M2). The signal and idler photons were separated from each other by a
polarizing beam splitter (PBS).
Vertically polarized photons (signal) are reflected by the PBS, and
diffracted by a grating (G) (1400 grooves/mm).
Horizontally polarized photons (idler) pass through the PBS and a
partially absorptive sample.

\begin{figure}[htbp]
    \begin{center}
     \includegraphics[scale=0.4]{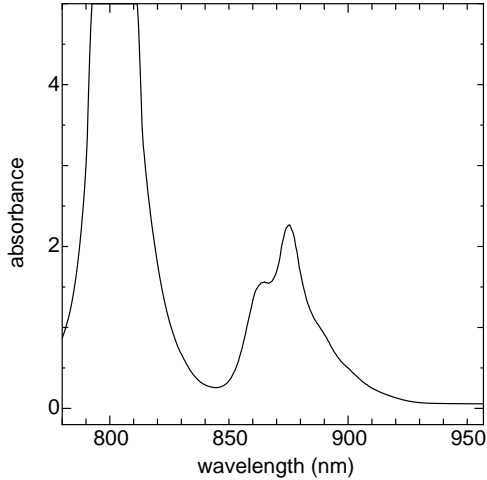}
    \end{center}
    \caption{Absorption spectrum of Nd$^{3+}$-doped
    glass.\label{data-Nd-absorbance}}
   \end{figure}

A 2.5-mm-thick plate of Nd$^{3+}$-doped glass was used as a
sample.
The main absorption transitions in Nd$^{3+}$ around the visible spectral
range are from $^4I_{9/2}$ to $^2G_{7/2}(^2G_{5/2})$, $^4F_{7/2}$,
$^4F_{5/2}$, and $^4F_{3/2}$, of which peak wavelengths are located at about
580, 750, 810, and 870~nm, respectively. The spectrum of signal light
is centered at 840~nm, and it overlaps with the absorption peaks at
810 and 870~nm (see Fig.~\ref{data-Nd-absorbance}), of which
transitions are studied in this paper. 

Both signal and idler photons are collimated into fibers by lenses
(L2,L3) and then detected by a set of two single-photon counting
modules (SPCM: EG\&G SPCM-AQR-14). Delaying the electric signal from the SPCM
detecting signal photons by a nanosecond delay unit (ORTEC 425A), the
coincidences are counted by a time-to-amplitude converter/single-channel
analyzer (ORTEC 567) followed by a computer-controlled
multichannel pulse-height analyzer (MCA:ORTEC TRUMP-PCI-2k).

\section{Theory}

In this paper, we discuss the application of the frequency-entanglement
of SPDC photon pairs to absorption spectroscopy. The state of generated SPDC photon
pairs can be written
as~\cite{coincident-frequencies-theory,coincident-frequencies-theory2,coincident-frequencies-theory3}
\begin{eqnarray}
|\psi\rangle \propto \int d\omega_s \int d\omega_i \tilde{E}(\omega_s+\omega_i) \Phi_L(\omega_s,\omega_i) |\omega_s\rangle_s |\omega_i\rangle_i,
\label{eq-state}
\end{eqnarray}
where $|\omega\rangle_j \equiv a^{\dagger}_j(\omega) |0\rangle$ is a
single-photon state whose frequency is $\omega$. $a^{\dagger}_j(\omega)$
is the photon creation operator for frequency $\omega$ photons. Here
$j=p,s,$ and $i$ indicate the pump, signal, and idler waves, respectively.
$\tilde{E}(\omega)$ is the Fourier transform of the classical
field of pump laser, and
\begin{eqnarray}
 \Phi_L(\omega_s,\omega_i) \propto \mathrm{sinc} \left[\Delta k(\omega_s,\omega_i)L/2\right] \equiv \frac{\sin(\Delta k(\omega_s,\omega_i)L/2)}{\Delta k(\omega_s,\omega_i)L/2}
\label{eq-phasematch}
\end{eqnarray}
is the phase-matching function~\cite{shg-phasematch} with the phase
mismatch parameter $\Delta k$ expressed as
\begin{eqnarray}
 \Delta k(\omega_s,\omega_i) = k_s(\omega_s)+k_i(\omega_i)-k_p(\omega_s+\omega_i).
\end{eqnarray}
As easily seen from Eq.~(\ref{eq-phasematch}),
thinner crystals are preferable for the generation of broadband SPDC photon
pairs. Generally, in our quantum experiments using SPDC photon pairs, thick
crystals as thick as 5~mm are usually used to obtain high SPDC conversion
efficiency. However in this paper broadband SPDC photon pairs are
indispensable for its spectroscopic measurement, thin crystal of 1-mm
thickness is used.

The wavelength resolution of this system is calculated as 4~nm using
$d_g\phi_f/2F$, where $d_g$, $\phi_f$, and $F$ are a gap of
grating grooves($1/1400$~mm), a fiber diameter ($125~\mu$m), and a
focal length of lens L2 (11~mm), respectively.
It is much narrower than the full width of the parametric fluorescence spectrum.

Since a cw laser is being used as a pump, the pump bandwidth is also
much narrower than the width of the parametric fluorescence spectrum.
It forces the sum of signal and
idler frequency to be equal to the pump frequency as
$\omega_p = \omega_s + \omega_i$
, and it entangles a SPDC photon pair in its frequency
as a signal photon at frequency $\omega_p/2-\omega$ is generated with an
idler photon at frequency $\omega_p/2+\omega$.

The wavelength resolution of the system and the pump bandwidth are so narrow
that, using some approximations, the absorption spectrum of the sample
can be easily calculated as~\cite{quantum-spectrum-filter,quantum-spectrum-filter2}
\begin{eqnarray}
 A(\omega^{'})=-\log_{10}\frac{R_c(\omega_p-\omega^{'})}{R_{c,\mathrm{sample}}(\omega_p-\omega^{'})},
\end{eqnarray}
where $R_{c,\mathrm{sample}}(\omega)$ is a coincidence counting rate
when the sample was placed in the idler path, and $R_c(\omega)$ is the
one without the sample.

\section{Results and Discussion}

\begin{figure}[htbp]
     \begin{center}
      \includegraphics[scale=0.4]{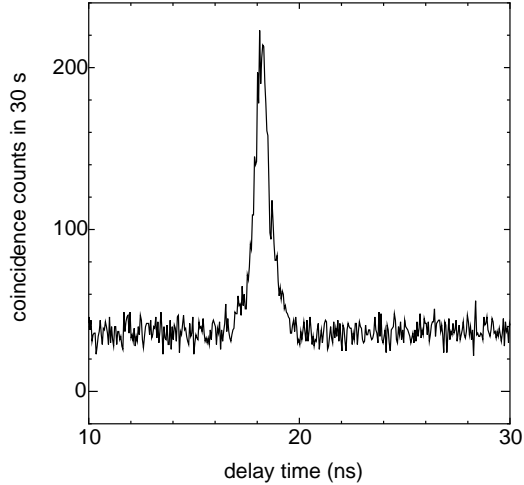}
     \end{center}
     \caption{Coincidence counts accumulated for 30 seconds when the
     grating angle is set to diffract center wavelength of signal
     light\label{data-coincidence-sample}}
    \end{figure}

Figure ~\ref{data-coincidence-sample} shows coincidence counts accumulated
for 30 s without a sample when the grating angle is set to
diffract the center wavelength of signal light, and the coincidence counts
have a peak at the delay time of 18~ns determined by the delay unit.

Coincidence counts averaged from 5 to 45~ns except between 14 and 22~ns delay is
used for the calculation of a background noise.
In this paper, coincidence sum counts is obtained by the integration of
the coincidence counts during the delay from 14 to 22~ns and by
subtracting the background noise.

\begin{figure}[htbp]
    \begin{center}
     \includegraphics[scale=0.4]{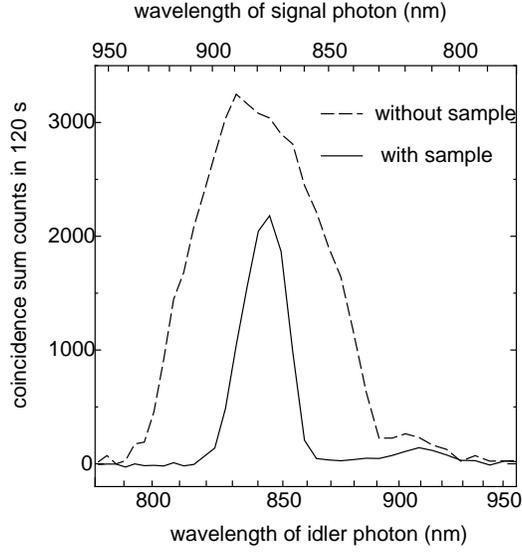}
    \end{center}
    \caption{Coincidence spectrum with the Nd$^{3+}$-doped glass
    sample, accumulated scanning a grating angle (solid line) and one
    without a sample (dashed line).\label{data-ndglass}}
   \end{figure}

Rotating the grating around the vertical axis crossing the
incident point of the idler beam, coincidence counts are accumulated for 120
s at each step without a sample. The dashed curve in
Fig.~\ref{data-ndglass} shows the wavelength dependence of the sum of
coincidence counts at each step (coincidence spectrum). The center
wavelengths of the signal and idler are centered at 883 and 840~nm,
respectively, and the full width at half maximums (FWHMs) are 63 and 69~nm, respectively.

Then, rotating the grating, coincidence spectrum was measured
with the Nd$^{3+}$-doped glass sample in the idler light path,
accumulating for 120 s at each step (see
Fig.~\ref{data-ndglass}). The wavelength calibration is performed using
a reflection, zeroth-order diffraction, from the grating. Beside the center
wavelength of the SPDC 
light, the glass sample has two absorption peaks at about 810 and 870~nm
(see Fig.~\ref{data-Nd-absorbance}), so the idler photons are absorbed
and the coincidence counts are reduced in the spectral ranges.

   \begin{figure}[htbp]
    \begin{center}
     \includegraphics[scale=0.4]{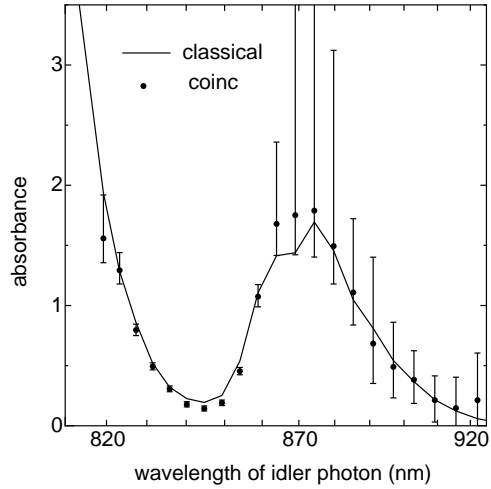}
    \end{center}
    \caption{The absorption spectrum calculated by dividing the
    coincidence spectrum without a sample by one with the sample. The
    solid line shows the absorption spectrum measured by a UV-visible-near-infrared
    spectrophotometer.\label{data-compareAbs}}
   \end{figure}

By dividing the coincidence spectrum without a sample by one with the
sample, the absorption spectrum of the sample is calculated, and
compared with one measured by a UV-visible-near-infrared spectrophotometer in
Fig.~\ref{data-compareAbs}. The absorption spectrum determined from the
ratio of the coincidence counts fits fairly well with the one measured by the conventional
spectrophotometer except in the spectral range from 860 to 880~nm,
where the sample absorbance is large and the amount of
transmitted photons is so small that the error tends to be substantially
large.

\section{Conclusion}

In conclusion, the absorption spectrum of the Nd$^{3+}$-doped glass
plate was measured using an frequency-entangled two-photon state
generated by spontaneous parametric down-conversion. This method is
performed without resolving the frequency of idler photons which
transmit through the sample. It is greatly advantageous in the case when
the transmitted photons are not easy to be resolved by their energy,
like in vacuum ultraviolet range or in space. The method using the
frequency entangled photon pairs has an advantage over the one using
classical method, when the spectroscopic property of a fragile sample is
analyzed in a spectral range where any low-noise photodetector is not
available, like in infrared range. Not to damage fragile samples, the
pump light power must be very low, but the photodetector for a infrared
range is usually so noisy that it is difficult to measure the
characteristics under such situation. However, using the
frequency-entangled photon pairs, one of the photon of each pair can be
used as a timing gate for the other photon of the pair resolved by its
energy. It increases the signal-to-noise ratio, so the measurement
becomes much easier than the case using the classical spectroscopy
apparatus.

A type-II BBO crystal is used for production of SPDC photon pairs
entangled in polarization. All photon pairs in the SPDC light
cones are collimated by a parabolic mirror, and travel to a beam
splitter. In this experiment, a polarizing beam splitter was used to
separate signal and idler photons from each other effectively. However
it destroys the polarization entanglement. Using a nonpolarizing beam
splitter in place of the polarizing beam splitter, the SPDC photon pairs
detected at photon counters have the polarization entanglement besides
the frequency entanglement, and it will enable WDM cryptography to be
implemented which has an ability to send much more information than the
case not using the frequency entanglement.

We thank Dr. Haibo Wang and Tomoyuki Horikiri for their valuable
discussion.



\end{document}